\documentclass[12pt]{article}
\usepackage{epsfig,psfrag}
\usepackage{graphicx,amsfonts,amsmath,amssymb}

\usepackage[usenames]{color}

\begin{document}

\begin{titlepage}

\begin{flushright}
\end{flushright}

\vskip 2cm
\begin{center}
{\Large \bf Reduction of the $\cal N$=8 BLG and $\cal N$=6 BL Theories to 2D Effective Field Theories}
\vskip 1.2cm
{\bf Paul Franche}

\vskip 0.4cm

{\it Rutherford Physics Building, McGill University, Montreal, QC H3A 2T8, Canada}

\vskip 0.2cm

{\tt franchep@hep.physics.mcgill.ca}

\vskip 1.5cm

\abstract{Starting from the three dimensional $\cal N$=8 BLG and $\cal N$=6 BL theories with CS level k=1, a dimensional reduction is performed on the M2-branes worldvolume. The IR limit of these 2D EFTs should represent the action of F1 strings in type IIA. The CS term reduces to a particular $\phi$F term and the scalars and fermions get ``mass'' terms. The reduced actions still rely on three-algebra, which we did not require to define precisely.}

\end{center}
\end{titlepage}

\section{Introduction}

In the past months, a lot of work have been done on the three dimensional superconformal gauge theory which were conjectured to described the worldvolume of N M2-branes. The first serious attempt has been done by Bagger and Lambert \cite{FirstBL, OriginalBL, ThirdBL} and Gustavsson \cite{Gustav1, Gustav2}. This theory, called BLG theory, has $\cal N$=8 supersymmetries, SO(8) R-symmetry, as expected on N M2-branes, and relies on an algebraic structure called 3-algebra.

The reduction of this N M2-branes theory to the ten dimensional type IIA string theory was done in \cite{Mukhi, M2D2Revisited} where they obtained the expected gauge theory living on a N D2-branes, ie 3 dimensional maximally supersymmetric YM. This was done using a novel Higgs mechanism which generated dynamics from the Chern-Simons (CS) term.

The $\cal N$=8 BLG theory worked on the assumption that the 3-algebra admitted a positive definite metric. This conditon was shown to be satisfied by only one choice of 3-algebra \cite{N8SuperCSThy} and was later relaxed to define Lorentzian 3-algebras \cite{M2D2Revisited, GMR, N8SuperGaugeThy, EMP, CS}. These Lorentzian 3-algebras were used in the compactification to D2-branes and the ghosts appearing there were taken care of in \cite{BLS, GRGVRV}.

In principle, M2-branes could also be reduced to F1 strings in type II string theory with a compactification along the direction of the M2-branes worldvolume. Looking for string actions coming from the reduction of M2-brane actions is the purpose of this paper. The original M2-M5 system then reduces to F1 between D4 or NS5 branes, depending on the embedding of the M5-branes. This was attempted in \cite{Scoopeurs}, but a detailed discussion was not presented therein.

Recently, another new 3 dimensional field theory with $\cal N$=6 supersymmetry and SU(4)$\times$U(1) R-symmetry was conjectured to represent the N M2-branes worldvolume \cite{ABJM}. Shortly after this, Bagger and Lambert also came with a new theory which had the same supersymmetry, R-symmetry and was based on a new complex 3-algebra structure \cite{NewBL}. With a particular choice of 3-algebra, this new BL theory was shown by the authors to be equivalent to the ABJM theory at k=1.

The paper is organised as follows.  In section 2.1, I review the k=1 $\cal N$=8 BLG theory. In section 2.2, I perform a dimensional reduction along the M2-brane worldvolume, i.e. ${\cal M}_2\times S^1$, to get an effective field theory in 2 dimensions which in the IR limit could represent a field theory on the F1 strings in type IIA.

In section 3.1, I review the k=1 $\cal N$=6 BL theory and in section 3.2, I again do a dimensional reduction on ${\cal M}_2\times S^1$ to get an effective fields theory in 2 dimensions which in the IR limit could also represent a field theory on the F1 strings in type IIA.

In both dimensional reductions, the CS action reduces to a particular $\phi$F term. However, the final two dimensional actions still rely on 3-algebras and the scalars and fermions get a ``mass'' term from $\tilde{\phi}$. I end with a brief discussion of the results.

All this work assumes that the 3-algebra admits a positive definite metric and hopes that the eventual ghosts could be taken care of.

Note added: While this work was in progress, \cite{Scoopeurs} appeared which has overlap with section 2.2 of this paper.

\section{The ${\cal N}$=8 Bagger-Lambert-Gustavsson Theory}

\subsection{The original 3D ${\cal N}$=8 BLG}

The BLG theory has $\cal N$=8 supersymmetry, SO(8) R-symmetry and is formulated in terms of real massless 3-algebra valued fields \cite{OriginalBL}. We take k=1 to simplify the calculations, but it should be easy to generalise.

First, let's review the BLG action
\begin{equation}
{\cal L}^{{\cal N}=8}_{3D} = -\frac{1}{2}(D_\mu X'^{aI})(D^\mu X'^I_{a}) + \frac{i}{2}\bar\psi'^a\Gamma^\mu D_\mu\psi'_a + \frac{i}{4}\bar\psi'_b\Gamma_{IJ} X'^I_c X'^J_d\psi'_af^{abcd} - V' + {\cal L}_{CS}^{3D} \label{OrigBLLagran}
\end{equation}

where
\begin{align}
V' = \frac{1}{12}f^{abcd}{f^{efg}}_dX'^I_aX'^J_bX'^K_cX'^I_eX'^J_fX'^K_g = \frac{1}{12}Tr([X'^I,X'^J,X'^K],[X'^I,X'^J,X'^K])\label{PotBLG}
\end{align}

and at k=1
\begin{equation}
{\cal L}_{CS}^{3D} = \frac{1}{2}\varepsilon^{\mu\nu\lambda}(f^{abcd}A'_{\mu ab}\partial_\nu A'_{\lambda cd} + \frac{2}{3}{f^{cda}}_gf^{efgb} A'_{\mu ab}A'_{\nu cd}A'_{\lambda ef}) \label{CSAction}\ .
\end{equation}

Here, $\mu , \nu , \lambda$ = 0,1,2 run along the internal dimensions of the M2, I,J = 1...8 are the transverse directions and a,b, ..., g are the 3-algebra generator indices. I gave all the 3D fields a prime (') because I'll drop it later for the rescaled 2D fields.

To allow the algebra to close, the structure constants must satisfy the following fundamental identity:
\begin{equation}
{f^{efg}}_d {f^{abc}}_g= {f^{efa}}_g {f^{bcg}}_d + {f^{efb}}_g {f^{cag}}_d + {f^{efc}}_g {f^{abg}}_d\ .
\end{equation}
Note that here the structure constants are real and totally antisymmetric.

The SUSY transformations in this theory are
\begin{align}
\delta X'^I_a &= i\bar{\epsilon}\Gamma^I\psi'_a\\
\delta\psi'_a &= D_\mu X'^I_a\Gamma^\mu \Gamma^I\epsilon - \frac{1}{6}X'^I_bX'^J_cX'^K_d{f^{bcd}}_a\Gamma^{IJK}\epsilon \\
\delta \tilde{A}'_\mu {^b}_a &= i\bar{\epsilon}\Gamma_\mu\Gamma_IX'^I_c\psi'_d{f^{cdb}}_a
\end{align}
where any the tilde represents contraction with the structure constant, i.e. ${\tilde A} {{}^a}_b = {f^{cda}}_b A_{cd}$. These transformations close on translations, gauge transformations and the following set of equations of motions:
\begin{align}
D^2X'^I_a - \frac{i}{2}\bar\psi'_c{\Gamma^I}_J X'^J_d\psi'_b{f^{cdb}}_a - \frac{\partial V'}{\partial X'^{Ia}} &= 0\label{eom1}\\
\Gamma^\mu D_\mu \psi'_a + \frac{1}{2}\Gamma_{IJ}X'^I_cX'^J_d\psi'_b{f^{cdb}}_a &= 0\label{eom2}\\
{\tilde F'} {{_{\mu \nu}}^a}_b + \varepsilon_{\mu \nu \lambda}(X'^J_c D^\lambda X'^J_d + \frac{i}{2}\bar\psi'_c \Gamma^\lambda\psi'_d){f^{cda}}_b &=0\label{eom3}\ .
\end{align}
These equation of motions led to the lagrangian \eqref{OrigBLLagran}.\\
The $\tilde{F'} {^{\mu \nu\ a}}_{b}$ and the covariant derivatives are defined as
\begin{align}
-\tilde{F'} {^{\mu \nu\ a}}_{b} &= \partial^\mu \tilde{A'} {^{\nu \ a}}_b - \partial^\nu \tilde{A'} {^{\mu \ a}}_b + \tilde{A'} {^{\mu \ a}}_{c}\tilde{A'} {^{\nu \ c}}_{b} - \tilde{A'} {^{\nu \ a}}_{c}\tilde{A'} {^{\mu \ c}}_{b}\\
(D_\mu X^I)_a &= \partial_\mu X^I_a - \tilde{A} {{_\mu}^b}_aX^I_b\ .
\end{align}

\subsection{Reduction of the ${\cal N}$=8 BLG to 2D}

Now, let's do a dimensional reduction along one internal direction of the M2. In order to do so, I reduce my $A'_\mu$ from 3 to 2 dimensions and I isolate the dependance on the extra $A'_2$. So, we are left with 2 $A'_{iab}$ and a scalar $A'_{2ab} = \phi'_{ab}$, where $i,j = 0,1$ are the remaining 2D indices, a,b,...,g indices are the 3-algebra indices and I choose fields to be independent of $x^2$. I integrate over the $x^2$ circle of circumference R, i.e. $\int dx^2=R$. More precisely\footnote{In this paper, I am not considering the effects of non-zero gauge fields and Wilson lines. I will leave it for future work.}, I take
\begin{equation}
\partial_2A'_\mu = \partial_2X'^I_a = \partial_2\psi'_a = 0
\end{equation}
\begin{equation}
g_{\mu \nu} = \left(
\begin{array}{ccc}
g_{ij} & 0 \\
0 & 1
\end{array} \right) \label{RedMetric}
\end{equation}
and I rescale  the 3D fields to 2D fields with canonical kinetic and CS terms:
\begin{align}
X^I = R^{\frac{1}{2}}X'^I\qquad
\psi = R^{\frac{1}{2}}\psi'\qquad
A_{i} = A'_{i}\qquad
\phi_{} = RA'_{2}\ .
\end{align}

Applying this to the original $\cal N$=8 theory and after some tedious but simple calculations, the lagrangian becomes
\begin{align}
{\cal L}^{{\cal N}=8}_{2D} = -&\frac{1}{2}(D_i X^{aI})(D^i X^I_{a}) - \frac{1}{2R^2}\tilde{\phi}^{ab}\tilde{\phi} {_a}^cX^I_bX^I_c\nonumber\\
&+ \frac{i}{2}\bar\psi^a\Gamma^i D_i\psi_a - \frac{i}{2R}\bar{\psi}^a\Gamma^2\tilde{\phi} {_a}^b\psi_b\label{reduceBLaction}\\
&+ \frac{i}{4R}\bar\psi_b\Gamma_{IJ} X^I_c X^J_d\psi_af^{abcd} - V + {\cal L}_{CS}^{2D}\nonumber
\end{align}

where
\begin{align}
V = \frac{1}{12R^2}f^{abcd}{f^{efg}}_dX^I_aX^J_bX^K_cX^I_eX^J_fX^K_g = \frac{1}{12R^2}Tr([X^I,X^J,X^K],[X^I,X^J,X^K])\ .
\end{align}

The reduction of the CS action is done in the following way:
\begin{align}
{\cal L}_{CS}^{2D} &=\frac{1}{2}f^{abcd}(\phi_{ab}\partial^iA^j_{cd} + A^i_{ab}\partial^j\phi_{cd})\varepsilon_{ij}\  \big(-\frac{1}{2}\partial^j(\phi_{ab}A^i_{cd}\varepsilon_{ij}f^{abcd})\big) \nonumber \\
&\qquad +\frac{2}{2\cdot 3}{f^{cda}}_g f^{efgb}\big(\phi_{ab}A^i_{cd}A^j_{ef} - A^i_{ab}\phi_{cd}A^j_{ef} + A^i_{ab}A^j_{cd}\phi_{ef}\big)\varepsilon_{ij}\\
&= \big(\phi_{ab}\partial^i\tilde{A}^{jab}+\phi_{ab}\tilde{A} {^{ia}}_g \tilde{A} ^{jgb}\big)\varepsilon_{ij}\\
&=-\frac{1}{2}\phi_{ab}\tilde{F} ^{ijab}\varepsilon_{ij} \label{FinalCS}
\end{align}
where I added a total derivative in the first line and I used the following equality which follows from the fundamental identity:
\begin{equation}
\phi_{ba}\tilde{A} {^{ia}}_g\tilde{A} ^{jgb}\varepsilon_{ij} = \tilde{\phi}^{fg}\tilde{A} {{^i}_g}^e {A^j}_{ef}\varepsilon_{ij}\ . \label{IdenDeFund}
\end{equation}

The equations of motions \eqref{eom1} and \eqref{eom2} now become
\begin{align}
D^2X^I_a +\frac{\tilde{\phi} {^c}_a \tilde{\phi} {^b}_c}{R^2} X^I_b
- \frac{i}{2R}\bar\psi_c{\Gamma^I}_J X^J_d\psi_b{f^{cdb}}_a - \frac{\partial V}{\partial X^{Ia}} &= 0\\\
\Gamma^iD_i\psi_a + \frac{1}{R}{f^{cdb}}_a\big(\frac{1}{2}\Gamma_{IJ}X^I_cX^J_d-\phi_{cd}\Gamma^2\big)\psi_b=0
\end{align}
and the $\tilde{F} {^{ab}_{\mu \nu}}$ equations of motion is now split in two equations, one for $F_{ij}$ and the other for $F_{2i}$:
\begin{align}
{\tilde F} {{_{ij}}^a}_b + \varepsilon_{ij}{f^{cda}}_b(-\frac{1}{R^2}\tilde{\phi} {^e}_dX^I_eX^I_c + \frac{i}{2R}\bar\psi_c \Gamma^2\psi_d) &=0\\
D_i\tilde{\phi} {^a}_b + \varepsilon_{ij}{f^{cda}}_b(X^I_c D^j X^I_d + \frac{i}{2}\bar\psi_c \Gamma^j\psi_d) &=0\ .
\end{align}

The SUSY transformations expand as
\begin{align}
\delta X^I_a &= i\bar{\epsilon}\Gamma^I\psi_a \label{susy1}\\
\delta\psi_a &= D_i X^I_a\Gamma^i \Gamma^I\epsilon -\frac{1}{R} \tilde{\phi} {^b}_a X^I_b\Gamma^2\Gamma^I\epsilon - \frac{1}{6R}X^I_bX^J_cX^K_d{f^{bcd}}_a\Gamma^{IJK}\epsilon\\
\delta \tilde{A}_i {^b}_a &= \frac{i}{R}\bar{\epsilon}\Gamma_i\Gamma_IX^I_c\psi_d{f^{cdb}}_a\\
\delta \tilde{\phi} {^b}_a &= i\bar{\epsilon}\Gamma_2\Gamma_IX^I_c\psi_d{f^{cdb}}_a\label{susy4}\ .
\end{align}

Thus we find a simple action in 2D given by~\eqref{reduceBLaction} which may have some connection to F1 strings in type IIA. We will discuss more about this later. Next we go to the ${\cal N}$=6 new BL theory.

\section{The ${\cal N}$=6 new Bagger-Lambert theory}

\subsection{The original 3D ${\cal N}$=6 BL theory}

The new BL theory has $\cal N$=6 supersymmetry, SU(4)$\times$U(1) R-symmetry and is formulated in terms of complex massless 3-algebra valued fields. Here too, we take k=1 to simplify the calculations, but it should be easy to generalise for arbitrary k.

An important difference here with the BLG theory is that $f ^{abcd}$ is no longer assumed to be real and totally antisymmetric, but rather that it satisfies
\begin{equation}
f^{abcd} = -f^{bacd} = -f^{abdc} = f^{*cdab} \label{Contraintesf}\ .
\end{equation}
This means that we have to be more careful with 3-algebra indices. Also, the fundamental identity (which is chosen to ensure closure of the algebra) is changed accordingly. It becomes
\begin{equation}
 {f^{efg}}_b {f^{cba}}_d + {f^{fea}}_b {f^{cbg}}_d + {f^{*gaf}}_b {f^{ceb}}_d + {f^{*age}}_b {f^{cfb}}_d=0
\end{equation}
which is equivalent to
\begin{equation}
{f^{cef}}_gf^{gbad} - {f^{cba}}_gf^{gefd} = {f^{bcd}}_gf^{gefa} - {f^{bef}}_gf^{gcda}\ . \label{fund2ndvers}
\end{equation}
The contraction of the algebra indices is now defined as:
\begin{equation}
\tilde{A} {{_\mu}^c}_d = {f^{cba}}_d A_{\mu ab}\ .\footnote{Note the sign difference with the previous definition in BLG. This is to the origin of the sign difference between equations~\eqref{FinalCS} and~\eqref{CSNewBL}}
\end{equation}
As mentioned in \cite{NewBL}, $\tilde{A} _\mu ^{ab}$ is antihermitian, i.e. $(\tilde{A} _\mu ^{ab})^* = -\tilde{A} _\mu ^{ba}$. In order for this to be true and for the action to be real, $A _\mu ^{ab}$ also has to be antihermitian. Note that in the $\cal N$=8 BLG theory we did not need any such assumption since the symmetric part was eliminated by contraction with f, but here antihermiticity has to be imposed from the start.

The lagrangian in this theory is
\begin{align}
{\cal L} _{3D}^{{\cal N}=6} = &-D_\mu Z'^A_aD^\mu \bar{Z'}^a_A - i\bar{\psi'}^A_a\gamma^\mu D_\mu \psi'^a_A - V' +{\cal L}_{CS}^{3D}\nonumber\\
&-if^{abcd}\bar{\psi'}^A_d\psi'_{Aa}Z'^B_b\bar{Z'}_{Bc} + 2if^{abcd}\bar{\psi'}^A_d\psi'_{Ba}Z'^B_b\bar{Z'}_{Ac}\\
&+ \frac{i}{2}\varepsilon_{ABCD}f^{abcd}\bar{\psi'}^A_d\psi'^B_cZ'^C_aZ'^D_b - \frac{i}{2}\varepsilon^{ABCD}f^{cdab}\bar{\psi'}_{Ac}\psi'_{Bd}\bar{Z'}_{Ca}\bar{Z'}_{Db}\nonumber
\end{align}
where, ${\cal L}^{3D}_{CS}$ at k=1 is given by
\begin{equation}
{\cal L}_{CS}^{3D} = \frac{1}{2}\varepsilon^{\mu\nu\lambda}(f^{abcd}A'_{\mu cb}\partial_\nu A'_{\lambda da} + \frac{2}{3}{f^{acd}}_gf^{gefb} A'_{\mu ba}A'_{\nu dc}A'_{\lambda fe})\ .\footnote{Note that two indices are reversed from the first formulation in~\cite{NewBL}. Private discussions with J. Bagger confirmed this is the correct formulation.}
\end{equation}

The scalar potential is
\begin{equation}
V' = \frac{2}{3}\Upsilon'^{CD}_{Bd}\bar{\Upsilon'}^{Bd}_{CD}\ . \label{PotNewBL}
\end{equation}
where
\begin{equation}
\Upsilon'^{CD}_{Bd} = {f^{abc}}_dZ'^C_aZ'^D_b\bar{Z'}_{Bc} - \frac{1}{2}{f^{abc}}_d\delta^C_BZ'^E_aZ'^D_b\bar{Z'}_{Ec} + \frac{1}{2}{f^{abc}}_d\delta^D_BZ'^E_aZ'^C_b\bar{Z'}_{Ec}\ .
\end{equation}

The SUSY transformations in this theory are
\begin{align}
\delta Z'^A_d &= i\bar{\epsilon}^{AB}\psi'_{Bd}\\
\delta \psi'_{Bd} &= \gamma^\mu D_\mu Z'^A_d\epsilon_{AB} + {f^{abc}}_dZ'^C_aZ'^A_b\bar{Z'}_{Cc}\epsilon_{AB} + {f^{abc}}_d Z'^C_aZ'^D_b\bar{Z'}_{Bc}\epsilon_{CD}\\
\delta\tilde{A'} {{_\mu}^c}_d &= -i\bar{\epsilon}_{AB}\gamma_\mu Z'^A_a\psi'^B_b{f^{cab}}_d + i\bar{\epsilon}^{AB}\gamma_\mu \bar{Z'}_{Ab}\psi'_{Ba}{f^{cba}}_d\ .
\end{align}
These transformations close on translations, gauge transformations and the following set of equations of motions:
\begin{align}
0 &= \gamma^\mu D_\mu \psi'_{Cd} + {f^{abc}}_d\psi'_{Ca}Z'^D_b\bar{Z'}_{Dc} - 2{f^{abc}}_d\psi'_{Da}Z'^D_b\bar{Z'}_{Cc} - \epsilon_{CDEF}{f^{abc}}_d\psi'^D_cZ'^E_aZ'^F_b\label{Psieom}\\
0 &= \tilde{F'} {{_{\mu \nu}}^c}_d + \varepsilon_{\mu \nu \lambda}\big( (D^\lambda Z'^A_a)\bar{Z'}_{Ab}- Z'^A_a(D^\lambda\bar{Z'}_{Ab}) - i\bar{\psi'}^A_b\gamma^\lambda\psi'_{Aa}\big){f^{cab}}_d\ .
\end{align}
The equation of motion of the Z scalars should be found taking the supersymetric variation of the $\psi$ equation of motion~\eqref{Psieom}.
The covariant derivative is defined this time as:
\begin{equation}
D_\mu Z^A_d = \partial_\mu Z^A_d - \tilde{A} {{_\mu}^c}_d Z^A_c, \qquad D_\mu \bar{Z}_{Ad} = \partial_\mu \bar{Z}_{Ad} + \tilde{A} {{_\mu}^c}_d \bar{Z}_{Ac}\ .
\end{equation}

\subsection{Reduction of the ${\cal N}$=6 BL to 2D}

Now, I can do a similar dimensional reduction as before. I choose every fields to be independent of $x^2$, use the metric \eqref{RedMetric}, rename $A_2$ by $\phi$ and I rescale the 3D fields to 2D fields with canonical kinetic and CS terms in the following way:
\begin{align}
Z^A = R^{\frac{1}{2}}Z'^A\qquad
\psi_A = R^{\frac{1}{2}}\psi_A'\qquad
A_{i} = A'_{i}\qquad
\phi_{} = RA'_{2}\ .
\end{align}

Applying this to the original ${\cal N}$=6 theory, after some long and tedious calculations, the lagrangian becomes
\begin{align}
{\cal L} _{2D}^{{\cal N}=6} = &-D_\mu Z^A_aD^\mu \bar{Z}^a_A - \frac{1}{R^2}\tilde{\phi}{^c}_a\tilde{\phi}^{ea}Z^A_c\bar{Z}_{Ae} - V +{\cal L}^{2D}_{CS}\nonumber\\
&-i\bar{\psi}^A_a\gamma^\mu D_\mu \psi^a_A + \frac{i}{R}{\phi}^{ac}\bar{\psi}^A_a\gamma^2\psi_{Ac}\nonumber\\
&-\frac{i}{R}f^{abcd}\bar{\psi}^A_d\psi_{Aa}Z^B_b\bar{Z}_{Bc} + \frac{2i}{R}f^{abcd}\bar{\psi}^A_d\psi_{Ba}Z^B_b\bar{Z}_{Ac}\label{reduceNewBLaction}\\
&+ \frac{i}{2R}\varepsilon_{ABCD}f^{abcd}\bar{\psi}^A_d\psi^B_cZ^C_aZ^D_b - \frac{i}{2R}\varepsilon^{ABCD}f^{cdab}\bar{\psi}_{Ac}\psi_{Bd}\bar{Z}_{Ca}\bar{Z}_{Db}\nonumber
\end{align}
where
\begin{equation}
V = \frac{2}{3R^2}\Upsilon^{CD}_{Bd}\bar{\Upsilon}^{Bd}_{CD}
\end{equation}
and $\Upsilon^{CD}_{Bd}$ are defined now with rescaled fields as:
\begin{equation}
\Upsilon^{CD}_{Bd} = {f^{abc}}_dZ^C_aZ^D_b\bar{Z}_{Bc} - \frac{1}{2}{f^{abc}}_d\delta^C_BZ^E_aZ^D_b\bar{Z}_{Ec} + \frac{1}{2}{f^{abc}}_d\delta^D_BZ^E_aZ^C_b\bar{Z}_{Ec}\ .
\end{equation}
The CS action in this theory reduces to
\begin{align}
{\cal L}_{CS}^{2D} &= \frac{1}{2}\varepsilon^{ij}\big[f^{abcd}(\phi_{cb}\partial_i A_{jda} + A_{icb}\partial_j \phi_{da}) + \big(\frac{1}{2}\partial_i(f^{abcd}\phi_{cb}A_{jda})\big)\nonumber\nonumber\\
&\qquad + \frac{2}{3}{f^{acd}}_gf^{gefb} (\phi_{ba}A_{idc}A_{jfe} - A_{iba}\phi_{dc}A_{jfe} + A_{iba}A_{jdc}\phi_{fe})\big]\nonumber\\
&= \varepsilon^{ij}\big[\phi_{cb}\partial_i \tilde{A}_j^{bc} + \frac{1}{3}(\phi{^b}_a\tilde{A}{{_i}^a}_g\tilde{A}{{_j}^g}_b - A{{_i}^b}_a\tilde{\phi}{^a}_g\tilde{A}{{_j}^g}_b + A{{_i}^b}_a\tilde{A}{{_j}^a}_g\tilde{\phi}{^g}_b)\big]\nonumber\\
&= \varepsilon^{ij}\big[\phi_{cb}\partial_i \tilde{A}_j^{bc} + \phi_{ba}\tilde{A}{{_i}^a}_g\tilde{A}{_j}^{gb}\big]\nonumber\\
&= \frac{1}{2}\phi_{cb}\tilde{F}^{bc}_{ji}\varepsilon^{ij}\label{CSNewBL}
\end{align}
where I added a total derivative in the first line and I used the following identity which comes from the fundamental identity as expressed in~\eqref{fund2ndvers}:
\begin{equation}
\tilde{A}_\mu^{gd}A_{\nu dc}\tilde{A}{{_\lambda}^c}_g - \tilde{A}{{_\mu}^c}_gA_{\nu dc}\tilde{A}{_\lambda}^{gd} = A_{\mu ab}\tilde{A}_\nu{^{b}}_g \tilde{A}{{_\lambda}^{ga}} -A_{\mu ab}\tilde{A}_\nu^{ga} \tilde{A}{{_\lambda}^b}_g\ .
\end{equation}

The equation of motion of the fermion becomes
\begin{equation}
0 = \gamma^i D_i \psi_{Cd} + \frac{1}{R}\big(\tilde{\phi}{^c}_d\gamma^2\psi_{Cc} + {f^{abc}}_d\psi_{Ca}Z^D_b\bar{Z}_{Dc} - 2{f^{abc}}_d\psi_{Da}Z^D_b\bar{Z}_{Cc} - \epsilon_{CDEF}{f^{abc}}_d\psi^D_cZ^E_aZ^F_b\big)
\end{equation}
and the gauge field equations of motion split again in $F_{ij}$ and $F_{2i}$ in the following way:
\begin{align}
\tilde{F} {{_{ij}}^c}_d &= \varepsilon_{ij}\big(\frac{1}{R^2}(\tilde{\phi}{^e}_a Z^A_e\bar{Z}_{Ab} + Z^A_a \tilde{\phi}{^e}_b \bar{Z}_{Ae}) + \frac{i}{R}\bar{\psi}^A_b\gamma^2\psi_{Aa}\big){f^{cab}}_d\\
R\tilde{F} {{_{2i}}^c}_d = D_i\tilde{\phi}{^c}_d &= -\varepsilon_{ij}\big( (D^j Z^A_a)\bar{Z}_{Ab}- Z^A_a(D^j\bar{Z}_{Ab}) - i\bar{\psi}^A_b\gamma^j\psi_{Aa}\big){f^{cab}}_d\ .
\end{align}

The SUSY transformations become
\begin{align}
\delta Z^A_d &= i\epsilon^{AB}\psi_{Bd}\\
\delta \psi_{Bd} &= \gamma^i D_i Z^A_d\epsilon_{AB} +\frac{1}{R}\big(- \tilde{\phi}{^c}_dZ^A_c\gamma^2\epsilon_{AB} + {f^{abc}}_dZ^C_aZ^A_b\bar{Z}_{Cc}\epsilon_{AB} + {f^{abc}}_d Z^C_aZ^D_b\bar{Z}_{Bc}\epsilon_{CD}\big)\\
\delta\tilde{A} {{_i}^c}_d &= \frac{-i}{R}\big(\bar{\epsilon}_{AB}\gamma_i Z^A_a\psi^B_b{f^{cab}}_d - \bar{\epsilon}^{AB}\gamma_i \bar{Z}_{Ab}\psi_{Ba}{f^{cba}}_d\big)\\
\delta\tilde{\phi}{^c}_d &= -i\bar{\epsilon}_{AB}\gamma_2 Z^A_a\psi^B_b{f^{cab}}_d + i\bar{\epsilon}^{AB}\gamma_2\bar{Z}_{Ab}\psi_{Ba}{f^{cba}}_d\ .
\end{align}

Thus to conclude, by dimensional reduction here I get another action~\eqref{reduceNewBLaction} which should also be related to F1 strings in type IIA.

\section{Discussion}

The original 3D CFT described above are conjectured to described a M2-M5 system in M-theory. The IR limits, i.e. small R limit, of the compactifications presented above should be a reduction from M-theory to type IIA string theory. We could then be tempted to associate our 2D field theories to the worldvolume field theories of F1 strings between branes in type IIA (D4 or NS5, depending on the compactification of the M5). In the UV, i.e. large R limit, the actions~\eqref{reduceBLaction} and~\eqref{reduceNewBLaction} reduce to kinetic and CS terms, but this limit does not lead to IIA string theory. Looking at the $k\neq1$ case would be relevant since the real IR limit of the compactification implies different limits of k and R \cite{ABJM}.

Let us comment on the form of the 2D actions~\eqref{reduceBLaction} and~\eqref{reduceNewBLaction} that we are left with. The first notable difference between the compactification presented here and the one described in \cite{Mukhi, M2D2Revisited} is that the compactification direction is along the M2, ie perpendicular to the scalars. The associated R-symmetries in each theories are then fully preserved. The potentials~\eqref{PotBLG} and~\eqref{PotNewBL} and the Yukawa terms also remain unaltered by the compactification.

Secondly, its not clear to me what is the precise connection between the two 2D theories. However, it is important to notice that the full 3-algebras still applies at the 2D level and is not broken to a Lie algebra like in previous compactifications. This is related to the fact that we did not make any additional assumption about the 3-algebras. It would be interesting to see how other types of 3-algebra \cite{Classification3Alg} could be reduced to a 2D EFT.

Note also that $\tilde{\phi}$ introduces ``mass'' terms for the scalars and fermions in the action, almost like with a Higgs mechanism. Remember that $\phi$ and $A_\mu$ are still non-dynamical fields. The $\phi F$ term in the action is expected from the reduction of a CS term. We had to use the fundamental identity to reduce it to this form. What is particular here is the presence of the two 3-algebra indices on each of those two fields.

\begin{figure}[ht]
  \begin{center}
    \includegraphics[width=5cm]{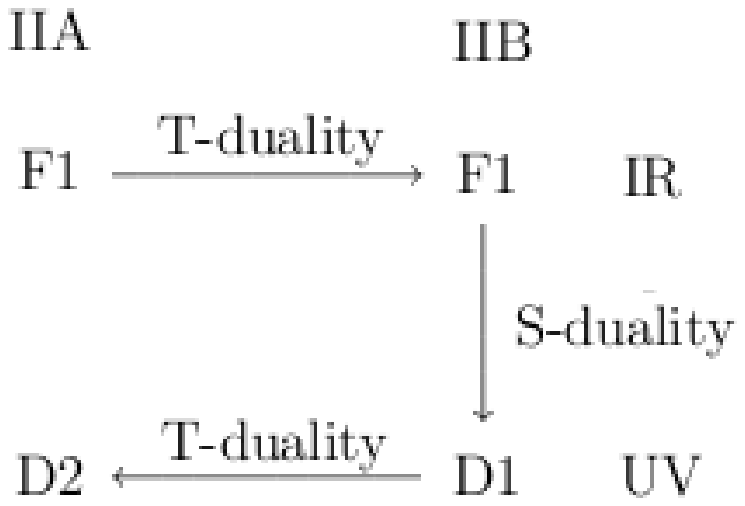}
    \caption{Set of dualities to map F1 strings to D2-branes}
    \label{SetDualities}
  \end{center}
\end{figure}

It is possible that our F1 string action could be related to D-brane actions. For example, it is possible that the $\frac{1}{R}$ terms in the actions could be mapped to the higher orders in $\alpha'$ of the D2-branes action shown in \cite{NewMukhi} through the set of dualities illustrated in fig.~\ref{SetDualities}. The large $\frac{1}{R}$ factors could be mapped to small $\alpha'$ factors because of the S-duality. Again, looking at the $k\neq1$ case would be relevant here.

\section{Acknowledgements}
The author would like to thank Jonathan Bagger and Keshav Dasgupta for useful discussions and comments on the first draft. The author also would like to thank Simon Caron-Huot, Andrew Frey, Rhiannon Gwyn and Alisha Wissanji for useful discussions and support. The work of the author was is supported by the Natural Sciences and Engineering Research Council of Canada (NSERC).

\section{Appendix: 2D chiral formulation of $\cal N$=8 BLG}

Due to the presence of the $\Gamma^2$ matrices in the reduced action, we could be tempted to expand the fermions into chiral components. Remember that in 2D (and 3D) \cite{Polchinski}

\begin{equation}
\Gamma^0 = 
\begin{bmatrix}
0 & 1\\
-1 & 0
\end{bmatrix}
, \qquad \Gamma^1 = 
\begin{bmatrix}
0 & 1\\
1 & 0
\end{bmatrix}
\qquad \text{and} \qquad \Gamma^2 = \Gamma^0 \cdot \Gamma^1 =
\begin{bmatrix}
1 & 0\\
0 & -1
\end{bmatrix}
\end{equation}
where $\Gamma^2$ can be used to define the projection matrix in 2D (like the $\gamma^5$ in 4D).

I can split the 11D fermion into its 3D chiral and anti-chiral and 8D parts:
\begin{equation}
\psi_{11D} = \left(
\begin{array}{ccc}
\psi^+_{8D}\\ \psi^-_{8D}
\end{array} \right)_{3D}
= \left(
\begin{array}{ccc}
\psi^+\\ \psi^-
\end{array} \right)
\quad \text{and} \quad \bar{\psi} = \left(
\begin{array}{ccc}
-\bar{\psi}^- & \bar{\psi}^+
\end{array} \right)\ .
\end{equation}
Writing this expansion explicity in~\eqref{reduceBLaction}, we get
\begin{align}
{\cal L}_{2D} = &-\frac{1}{2}(D_i X^{aI})(D^i X^I_{a}) - \frac{1}{2R^2}\tilde{\phi}^{ab}\tilde{\phi} {_a}^cX^I_bX^I_c - V + {\cal L}_{CS}^{2D}\label{LagrBL2D}\nonumber\\
&- \frac{i}{2}\big(\bar\psi^{+a}\Gamma^i D_0\psi^+_a - \bar\psi^{+a}\Gamma^i D_1\psi^+_a + \bar\psi^{-a}\Gamma^i D_0\psi_{-a} + \bar\psi^{-a}\Gamma^i D_1\psi_{-a}\big)\nonumber\\
&+ \frac{i}{2R}\tilde{\phi} {_a}^b\big(\bar{\psi}^{-a}\psi^+_b + \bar{\psi}^{+a}\psi^-_b\big)\\
&+ \frac{i}{4R}\big(-\bar\psi^-_b\Gamma_{IJ} \psi_a^+ + \bar\psi^+_b\Gamma_{IJ} \psi_a^- \big) X^I_c X^J_df^{abcd}\nonumber
\end{align}
and the SUSY transformations~\eqref{susy1} to~\eqref{susy4} become
\begin{align}
\delta X^I_a &= -i\bar{\epsilon}^-\Gamma^I\psi^+_a + i\bar{\epsilon}^+\Gamma^I\psi^-_a\\
\delta\left(
\begin{array}{ccc}
\psi_a^+\\
\psi_a^-
\end{array}\right)
&= D_0 X^I_a\Gamma^I\left(
\begin{array}{ccc}
\epsilon^-\\
-\epsilon^+
\end{array}\right)
 -\frac{1}{R} \tilde{\phi} {^b}_a X^I_b\Gamma^I\left(
\begin{array}{ccc}
\epsilon^-\\
\epsilon^+
\end{array}\right)
- \frac{1}{6R}X^I_bX^J_cX^K_d{f^{bcd}}_a\Gamma^{IJK}\left(
\begin{array}{ccc}
\epsilon^+\\
\epsilon^-
\end{array}\right)\\
\delta\left(
\begin{array}{ccc}
\tilde{A}_0 {^b}_a \\
\tilde{A}_1 {^b}_a
\end{array} \right)
&= \frac{i}{R}{f^{cdb}}_aX^I_c\left(
\begin{array}{ccc}
\bar{\epsilon}^+\Gamma_I\psi^+_d + \bar{\epsilon}^-\Gamma_I\psi^-_d\\
\bar{\epsilon}^+\Gamma_I\psi^+_d - \bar{\epsilon}^-\Gamma_I\psi^-_d
\end{array} \right)\\
\delta \tilde{\phi} {^b}_a &= -iX^I_c{f^{cdb}}_a\big(\bar{\epsilon}^-\Gamma_I\psi^+_d + \bar{\epsilon}^+\Gamma_I\psi^-_d\big)\ .
\end{align}

This could probably be done also with the reduced ${\cal N}$=6 BL lagrangian \eqref{reduceNewBLaction}.

\end{document}